\documentclass[conference]{IEEEtran}
\IEEEoverridecommandlockouts
\usepackage{cite}
\usepackage{amsmath,amssymb,amsfonts}
\usepackage{algorithmic}
\usepackage{graphicx}
\usepackage{textcomp}
\usepackage{xcolor}

\usepackage{upgreek}
\usepackage[hyphens]{url}
\usepackage{subcaption}
\usepackage{tikz}
\usetikzlibrary{shapes.geometric, arrows.meta, positioning}

\tikzstyle{block} = [rectangle, rounded corners, minimum width=3cm, minimum height=1cm,text centered, draw=black, fill=gray!10]
\tikzstyle{arrow} = [thick,->,>=stealth]

\def\BibTeX{{\rm B\kern-.05em{\sc i\kern-.025em b}\kern-.08em
    T\kern-.1667em\lower.7ex\hbox{E}\kern-.125emX}}
\begin{document}

\title{WIP: Turning Fake Chips into Learning Opportunities\\
 \thanks{$^{*}$ Indicates authors who contributed equally to this work.}
 \thanks{Haniye Mehraban was supported by the National Aeronautics and Space Administration (NASA) under Grant 80NSSC24M0161.}
}

\author{
  Haniye Mehraban$^{*}$, 
  Saad Azmeen-ur-Rahman$^{*}$, 
  John Hu \\
\textit{School of Electrical \& Computer Engineering}, Oklahoma State University, Stillwater, OK 74078, USA\\
\\

}

\maketitle

\begin{abstract}
This work-in-progress paper presents a case study in which counterfeit TL074 operational amplifiers, discovered in a junior-level electronics course, became the basis for a hands-on learning experience. Counterfeit integrated circuits (IC) are increasingly common, posing a significant threat to the integrity of undergraduate electronics laboratories. Instead of simply replacing the counterfeit components, we turned the issue into a teaching moment. Students engaged in hands-on diagnostics—measuring current, analyzing waveforms, and troubleshooting. By working with fake chip components, they gained deeper insight into analog circuits, supply chain security, and practical engineering.

\end{abstract}


\section{Introduction}

Hands-on electronics labs are foundational to undergraduate electrical and computer engineering (ECE) education, providing a critical bridge between abstract theory and real-world application. These lab environments often assume that the components used—especially standard ones like operational amplifiers—behave according to specifications. However, this assumption can break down when counterfeit integrated circuits (ICs) enter the educational supply chain, disrupting both the technical and pedagogical goals of lab-based learning.

Counterfeit ICs are a growing concern in hardware security and supply chain integrity. A counterfeit component is typically one that misrepresents its origin, manufacturer, or performance characteristics~\cite{sood_2011}. The rise in counterfeit electronics has attracted global attention due to its economic and operational consequences, with losses estimated in the trillions of dollars~\cite{kaenune_2013,guin_2014}. Although detection and prevention techniques such as watermarking, fingerprinting, split manufacturing~\cite{rostami_2014}, logic locking~\cite{cui_2022}, and EM-based analysis~\cite{stern_2020,ma_2019} exist, they often rely on expensive hardware and proprietary software that are not  accessible in consumer or educational settings. Moreover, most are designed to identify specific ICs that the system has been trained to recognize, making them impractical for general-purpose component screening in undergraduate labs.

At Oklahoma State University, a counterfeit chip incident became an unexpected pedagogical turning point. During the Fall 2024 semester, nearly every student in a junior-level electronics course received a fake TL074 operational amplifier in their lab kit. As circuits repeatedly failed to perform as expected, students began doubting their understanding, and teaching assistants (TAs) questioned the validity of their instructional materials. The issue was ultimately traced back to counterfeit ICs—a problem that had been growing over several semesters.

Rather than shielding students from the issue, we restructured the course to turn the incident into a hands-on learning opportunity.  Through measurement, analysis, and comparison, students gained a deeper understanding of component variability, supply chain security, and the essential role of testing in engineering.

This paper documents our investigation of counterfeit TL074 op-amps and characterizes their electrical behavior. We also detail the curriculum changes we implemented to turn a hardware failure into a meaningful learning experience. Through targeted lab adjustments and structured support, students were not only able to identify the problem but also learn from it—gaining practical insight into measurement, debugging, and the complexities of real-world circuit behavior. To guide this shift, we focused on two instructional goals:
\begin{enumerate}
    \item  Help students understand the root causes of unexpected circuit behavior. 
    \item  Use the situation to teach key engineering skills: measurement accuracy, component validation, troubleshooting, and critical thinking.
\end{enumerate}

\section{Background}

\subsection{TL074CN Opamp}

The TL07x series of opamps~\cite{tl074_ti,tl074_moto,tl074_hlf} is very popular among hobbyists, researchers, and professionals. It is used, for example, in audio applications~\cite{beware_2021,audioapp_1}, prototypes for biomedical research~\cite{bioapp_1,bioapp_4,bioapp_6,bioapp_9,bioapp_10}, didactic platforms for undergraduate teaching~\cite{eduapp_1}, agricultural research~\cite{agroapp_1}, and electronics research~\cite{eeapp_1,eeapp_2,phyapp_1}. However, while its availability and usage have become ubiquitous, so have its counterfeits~\cite{reddit_2024,beware_2021,bitten_2020,stack_2019}.

\subsection{Counterfeits of TL074}

Counterfeits of TL074 have been reported by a Canadian company that builds audio synthesizers~\cite{beware_2021}. They claim to have bought the ICs from Jameco Electronics, and noticed differences in the appearance of the package compared to a TL074 bought from Mouser. Others have reported purchasing similar counterfeits of TL074 from AliExpress~\cite{beware_2021,bitten_2020,stack_2019}, eBay~\cite{noel_2020}, and Tayda~\cite{bitten_2020}.

These fake parts have been reported with logos of, or similar to, Texas Instruments~(TI)~\cite{beware_2021,bitten_2020}, and ST Microelectronics~\cite{bitten_2020}, using different typefaces~\cite{bitten_2020}. Users have suggested checking for ``tinning" of IC pins as a sign of being a recycled part~\cite{bitten_2020}. Using acetone to wipe down the text on the package to check for fading has been proven inconclusive~\cite{beware_2021,noel_2020}. More effective electrical tests include resistance checks, building custom testing platforms, plugging into a system that uses the IC to observe change in operation~\cite{noel_2020}, conducting diode tests between each pair of pins, comparing amplifier outputs~\cite{beware_2021}, measuring slew rate (SR)~\cite{beware_2021,stack_2019}, and using opamp testers~\cite{reddit_2024}. One user opined that measuring the quiescent current may be enough to identify the genuine TL074 from the counterfeit~\cite{bitten_2020}.

\subsection {Use of TL074 in Our Instructional Lab}

\begin{figure}[htbp]
\centering
\includegraphics[width=0.34\textwidth]{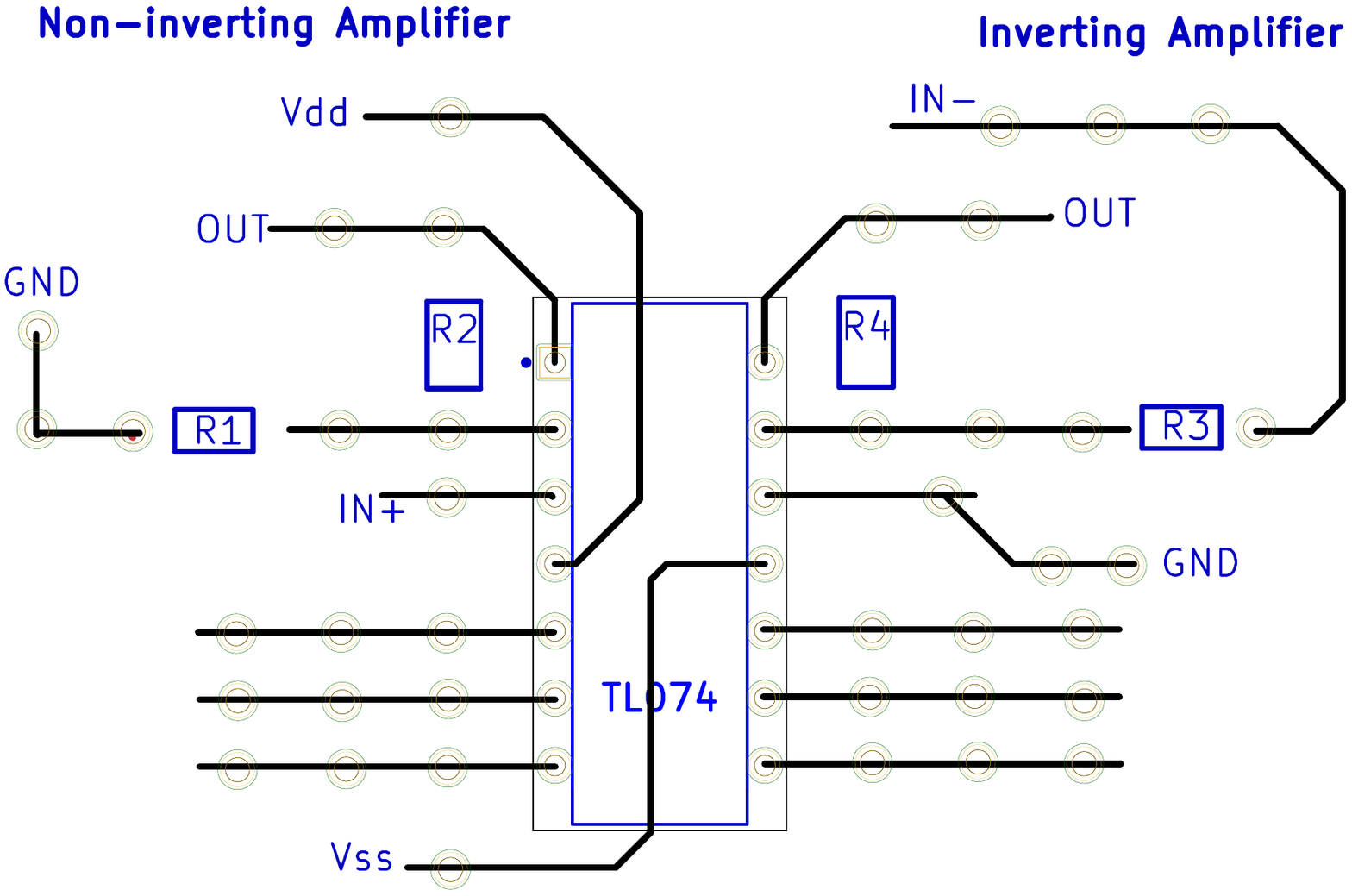}
\caption{Solder diagram of op-amp circuits on the custom PCB used in the lab.}
\label{fig:pcb_layout}
\end{figure}

We have used the TL074CN quad opamp, which is a 14-pin PDIP package~\cite{tl074_ti,tl074_moto}, in our instructional labs at Oklahoma State University for 7+ years to teach undergraduate students in their junior year. Students solder a TL074 and various passive components to build two amplifiers in their non-inverting and inverting configurations, with gains of 10 and 20, respectively. Students solder the TL074 and passive components onto a custom printed circuit boards (PCB)  to construct the circuits. The layout of this board is shown in Fig.~\ref{fig:pcb_layout}.

With a typical unity-gain bandwidth or gain-bandwidth product~(GBWP) of 3~to~4~MHz~\cite{tl074_ti,tl074_moto,tl074_hlf}, the amplifiers should therefore have 3~dB bandwidth~($f_{3dB}$) of 300+ and 150+~kHz, respectively. The experiment requires the frequency of a 1~$\mathrm{V_{p-p}}$ input sinusoidal signal to be swept from 50~Hz to 6~MHz to observe the gains and phases of the two amplifier configurations.

Since 2021, we have observed that some students had faulty ICs despite being bought from a reputable vendor. By Fall 2024, every lab section reported failures, which were eventually traced to counterfeit parts—primarily those labeled as TI-manufactured. This situation provided a unique challenge—and an opportunity to redesign the lab experience around real-world component variability, diagnostic methods, and supply chain uncertainty.

\section{Pedagogical Redesign and Curriculum Integration}
After identifying counterfeit TL074 op-amps as the cause of circuit failures, the instructional team redesigned the lab experience to transform the disruption into a structured learning opportunity. This section outlines the course context, the investigation into the counterfeit devices, and the development of a practical diagnostic method.

\subsection{ Course Context}

This instructional redesign took place in a junior-level undergraduate course on electronic devices and applications. The course introduces students to core semiconductor devices—including diodes, transistors, and operational amplifiers—with emphasis on circuit modeling, SPICE simulation, and amplifier design. Theoretical instruction is supported by the widely used textbook \textit{Microelectronic Circuit}s~\cite{sedra_smith}, and weekly lab sessions reinforce theoretical concepts through practical circuit implementation.

A central lab assignment involves designing and characterizing inverting and non-inverting amplifier circuits using the TL074CN op-amp. Students build these circuits on a custom PCB and perform frequency sweeps to observe gain and phase response, applying theoretical expectations to real-world circuit behavior.

\begin{figure}[!t]
    \centerline{\includegraphics[width=0.330\textwidth,trim={50pt 0pt 100pt 60pt},clip]{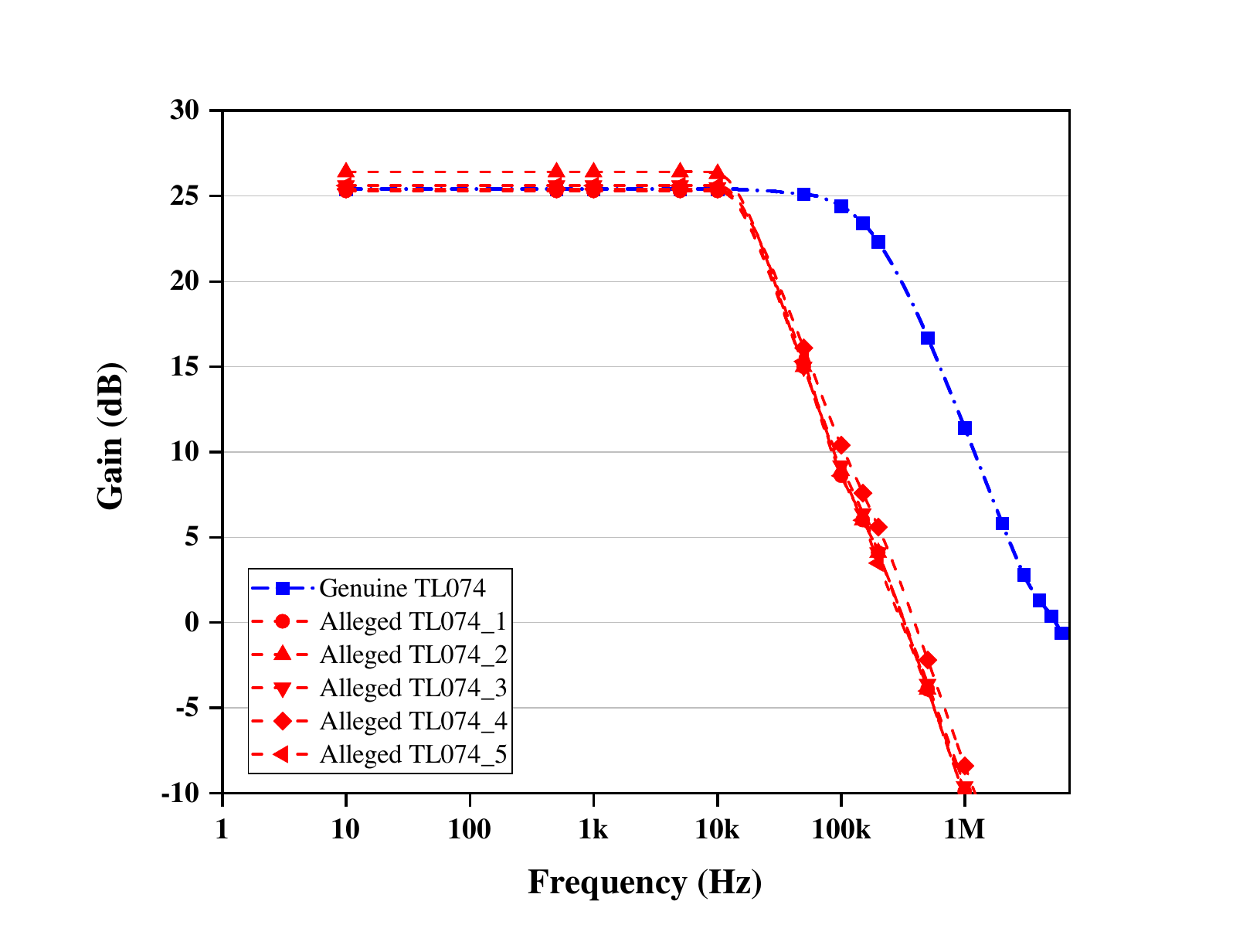}}
    \centerline{\includegraphics[width=0.330\textwidth,trim={50pt 30pt 100pt 60pt},clip]{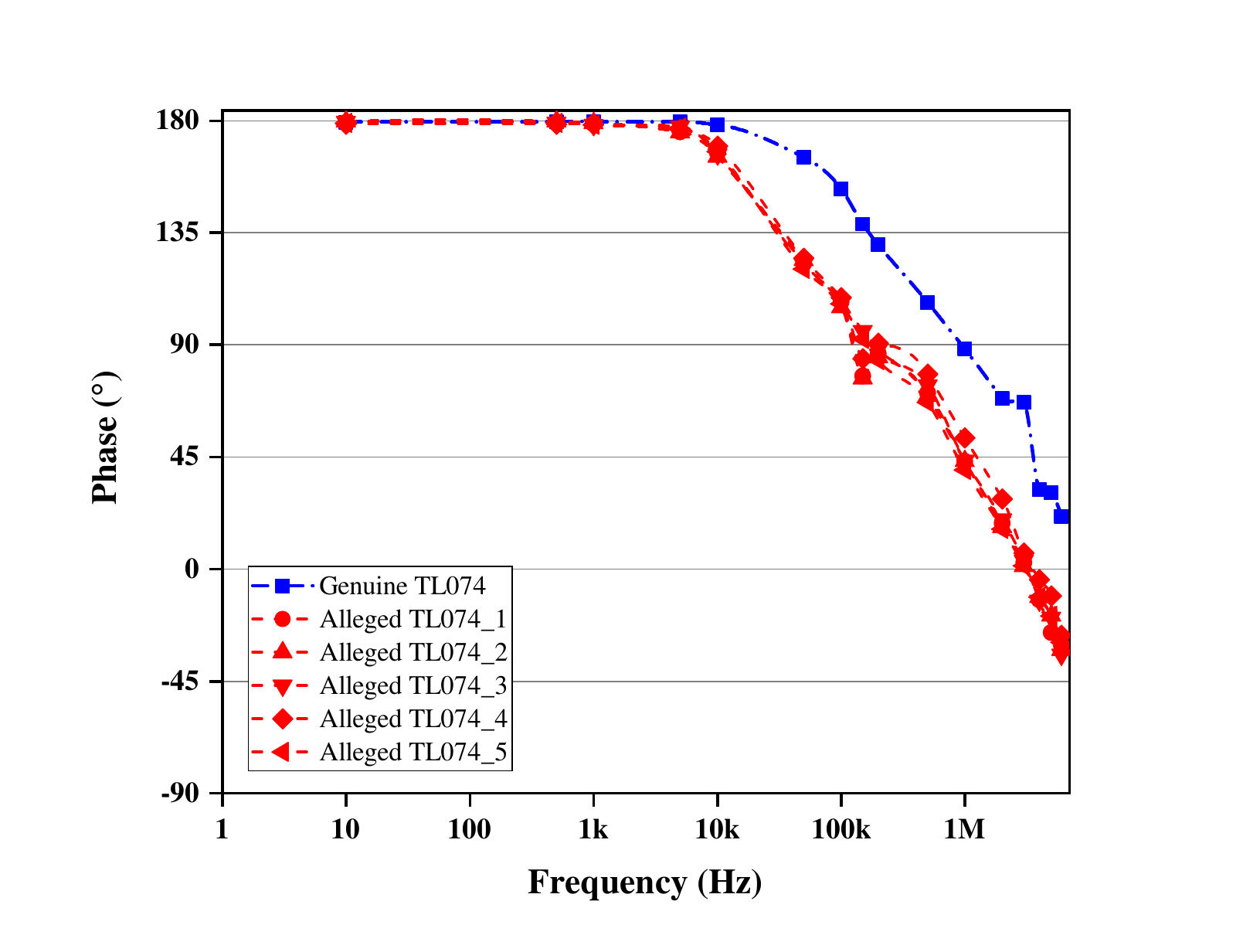}}
    \caption{Gain and phase response of genuine and alleged TL074 parts in inverting amplifier configuration with gain of 20~(26~dB).}
    \label{fig:bode}
\end{figure}

\begin{figure}[!t]
    
        \centerline{\includegraphics[width=0.31\textwidth]{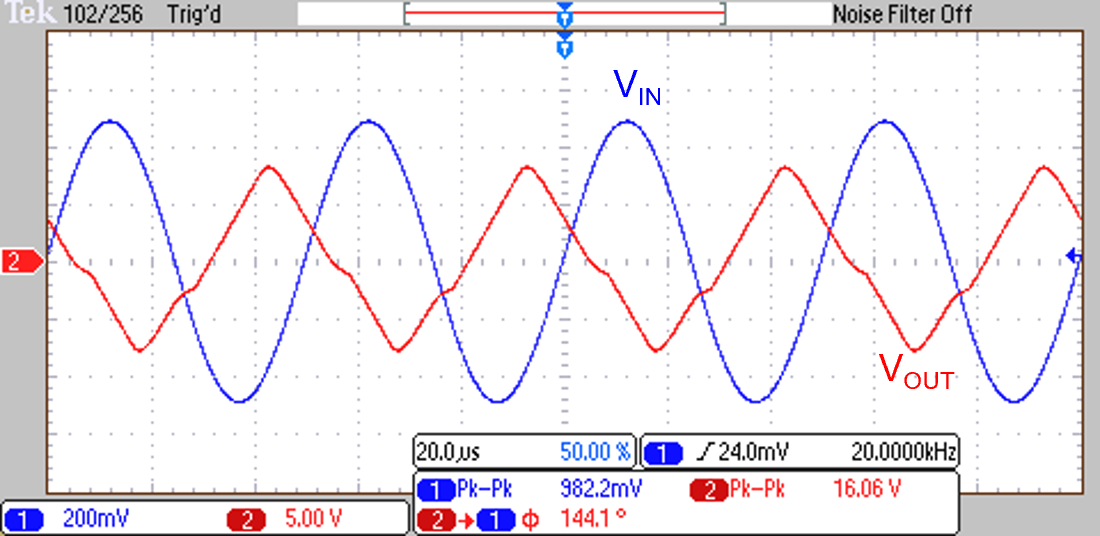}}
    \caption{Distortion of output signal due to alleged TL074 parts in inverting amplifier configuration at 20~kHz.}
    \label{fig:distorion}
\end{figure}

\subsection{Fake Chip Characterization}
To understand the root cause of the unexpected circuit behavior observed in the lab, we conducted two sets of electrical tests on both genuine and suspected counterfeit TL074CN op-amps.

In the first set, seven student PCBs, including one genuine Motorola and six alleged TI parts, were tested in the inverting amplifier configuration with a gain of 20.

All ICs were supplied with $\pm 15$~V and a 1~$\mathrm{V_{p-p}}$ input sinusoidal signal from 10~Hz to 6~MHz. The results in Fig.~\ref{fig:bode} show that the genuine Motorola TL074CN opamp produces the desired results with a voltage gain of 20~(26~dB), $f_{3dB}$ of 197~kHz with $45^{\circ}$ phase shift, and GBWP of 5.2~MHz. In contrast, the counterfeit opamps produced the same results irrespective of the manufacturer or date code. While the DC gain was 26~dB, the gain sharply dropped after 10~kHz, resulting in $f_{3dB}$ of only 22~kHz, and GBPW of 320~kHz. The waveform in Fig.~\ref{fig:distorion} shows how the output of the alleged TI opamp is distorted and looks more like a ramp signal at 20~kHz, caused by the limited SR of the counterfeit part. Closer observation showed that the distortion starts at approximately 200~Hz in the inverting amplifier configuration.
 
\begin{figure}
    \centering
    \begin{subfigure}{0.6\columnwidth}
        \includegraphics[width=\linewidth]{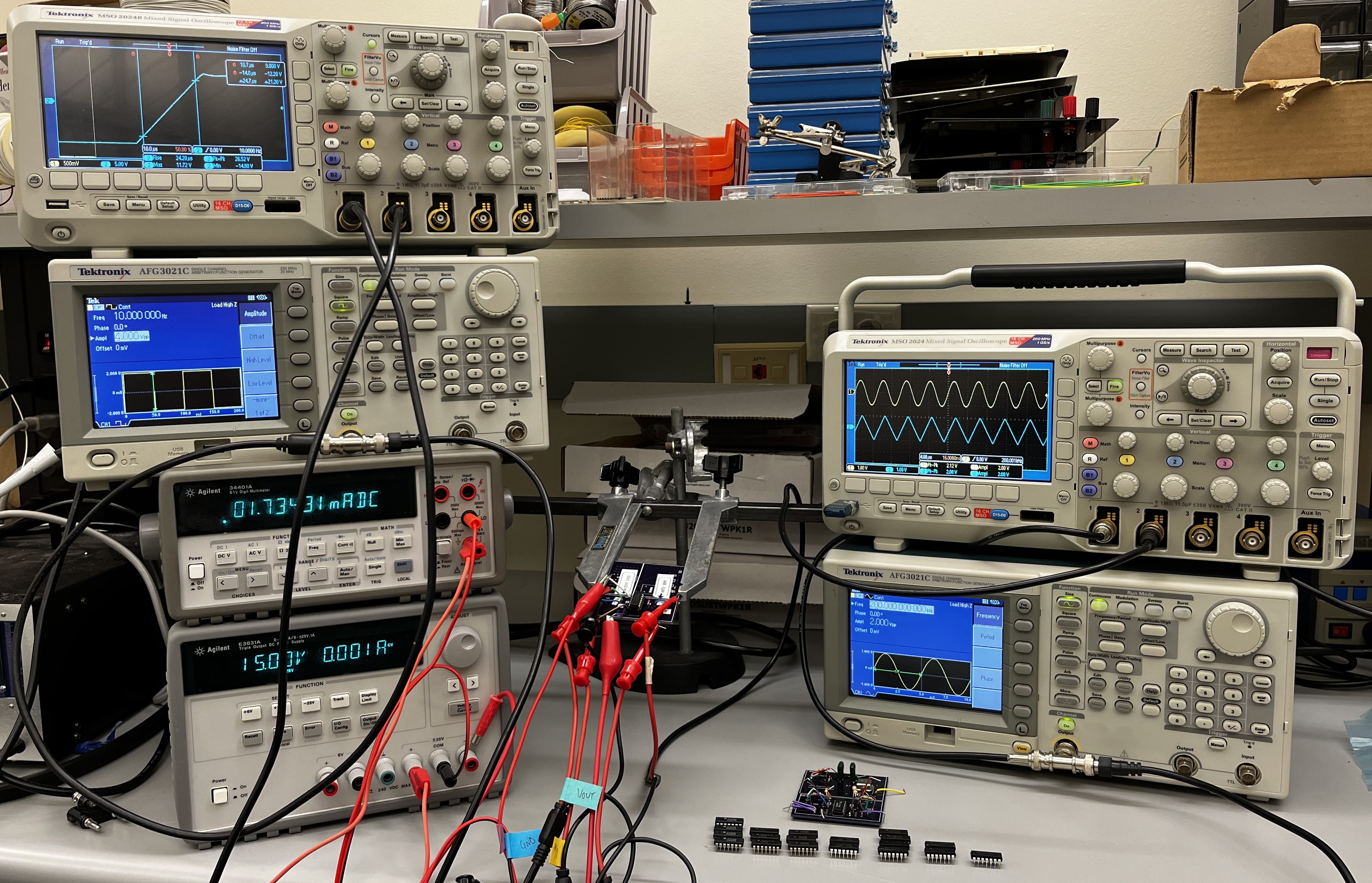}
        \caption{}
        \label{fig:fullsetup}
    \end{subfigure}
    \hfill
    \begin{subfigure}{0.33\columnwidth}
        \includegraphics[width=\linewidth]{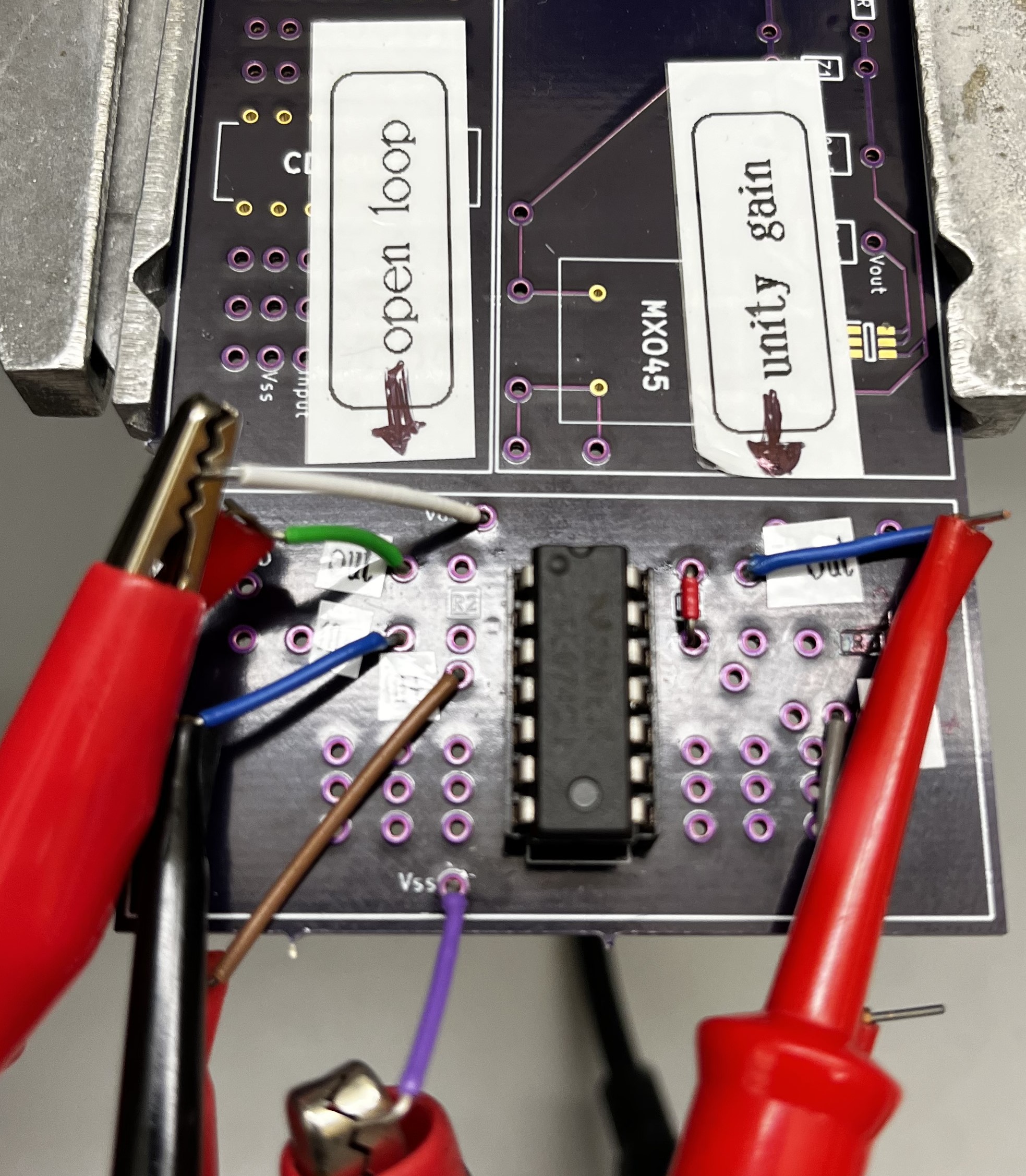}
        \caption{}
        \label{fig:pcb}
    \end{subfigure}
    \caption{(a) Experimental setup and (b) PCB constructed for open- and closed-loop electrical characterization of genuine and alleged TL074 parts.}
    \label{fig:setup}
\end{figure}

In the second set, 21 separate ICs that included five Motorola parts with two date codes, six TI parts with two date codes, two HFL parts with the same date code, and eight counterfeits from three manufacturers with logos resembling that of TI and with four different date codes were tested for various electrical characteristics of the TL074CN to classify real and fake parts. The maximum peak output voltage swing~($\mathrm{V_{OM}}$), GBWP, and SR, were measured.  The setup and PCB are shown in Fig.~\ref{fig:setup} -- an open-loop and a unity-gain amplifier configuration were set up using two of the opamps of the TL074CN IC. The opamps were powered by $\pm 15$~V supply. A 10~Hz 4~$\mathrm{V_{p-p}}$ input square wave was input to saturate the open-loop opamp to measure $\mathrm{V_{OM}}$ and SR and a 2~$\mathrm{V_{p-p}}$ input sinusoid was input to the unity-gain amplifier and the frequency swept to obtain GBWP.

\begin{figure}[!t]
    \centerline{\includegraphics[width=0.4\textwidth,trim={0pt 0pt 0pt 0pt},clip]{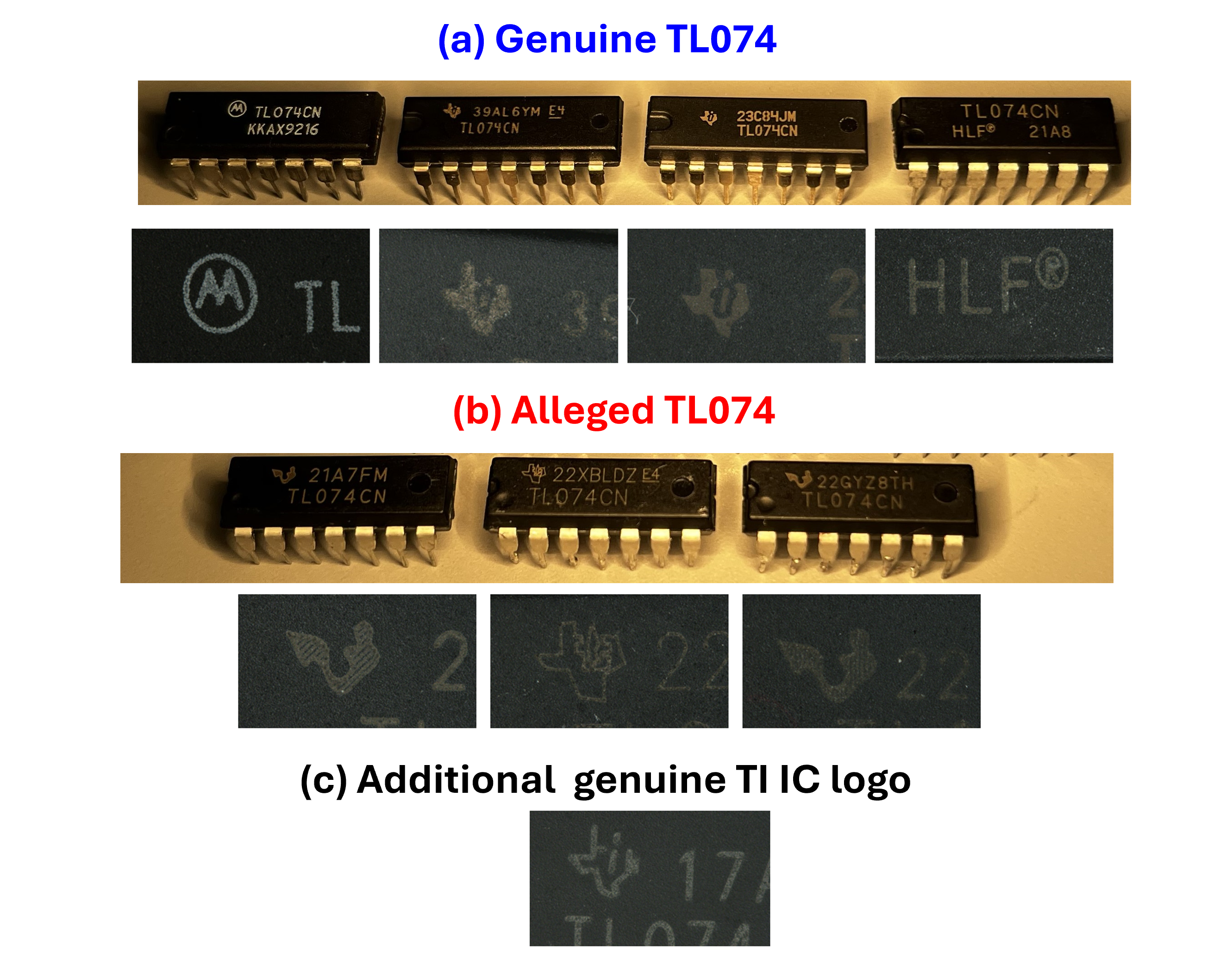}}
    \caption{Images of top and logos of (a)~genuine Motorola, TI, and HLF parts, (b)~alleged TI parts, and (c)~additional logo of a genuine TI part.}
    \label{fig:ic}
\end{figure}

\begin{figure}[!t]
    \centerline{\includegraphics[width=0.18\textwidth,trim={20pt 20pt 20pt 20pt},clip]{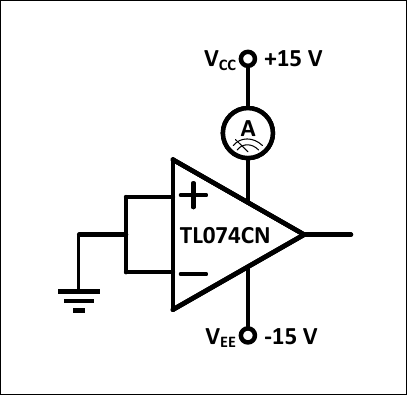}}
    \caption{Schematic of proposed $\mathrm{I_{CC}}$ detection method for TL074CN.}
    \label{fig:tl074schematic}
\end{figure}

In addition to electrical tests, a visual inspection of the TL074 packages was conducted. Photos of the different TL074CN packages used in our measurements are shown in Fig.~\ref{fig:ic}. Even to the trained eyes, it may be difficult to visually identify the counterfeits just by looking at the package and logo. Certainly, over the last 4 years, we have been unable to do so. The logos appear to be of TI but only under the right light conditions and angle, one may ``suggest'' that the logos look a little different. Once inspected under a microscope, the differences between genuine and fake logos are apparent. The additional genuine TI logo in Fig.~\ref{fig:ic}(c) is provided to show that manufacturers may have different logos across their products, and therefore, it would be very difficult to identify a genuine product from the logo simply with the naked eye.

These findings established a technical baseline that informed our instructional redesign.

\subsection{Proposed $\mathrm{I_{CC}}$ Detection Method}

To support student learning and circuit troubleshooting, we implemented a supply current~($\mathrm{I_{CC}}$) detection method, shown in Fig.~\ref{fig:tl074schematic}, to identify counterfeit TL074CN ICs based on their current consumption characteristics. Supplied with $\pm 15$~V and both inputs tied to ground, a genuine TL074CN part typically draws a current of 1.4--2.5~mA per amplifier in no-load condition~\cite{tl074_ti,tl074_moto,tl074_hlf}. We measured the supply current of the twenty-one ICs under these conditions. $\mathrm{I_{CC}}$ of the genuine TL074 parts were normally distributed with a mean and standard deviation of $\mu_{gen} = 1.89$~mA and $\sigma_{gen} = 0.135$, while those of the counterfeits were distributed with $\mu_{coun} = 0.42$~mA and $\sigma_{coun} = 0.023$. The upper-side limit~(USL) for $\mathrm{I_{CC}}$ is the maximum current of 2.5~mA specified in the datasheet. The lower-side limit~(LSL) can be taken from the middle point of the two distribution means with respect to their standard deviations, given by,

    \begin{equation}
        \label{eqn:statpower}
        k = \left | \frac{\mu_1-\mu_0}{\sigma_1+\sigma_0} \right |\mathrm{,}
    \end{equation}

\noindent which results in $k=9.3$, that is a LSL of 0.63~mA. The statistical power is greater than the six-sigma standard (or 5~ppm).

The smaller current drawn by the counterfeit parts also proves why they have a higher rise time and lower SR. Subsequently, at higher frequencies, the signal amplitude drops resulting in a lower GBWP, and the signal distorts and appears as a ramp signal.
Additionally, when one of the opamps was amplifying, an average increase in $\mathrm{I_{CC}}$ of 0.5~mA was observed with the genuine parts, whereas the counterfeit parts did not have a change in the current.  It is worth noting that $\mathrm{V_{OM}}$ cannot be used to determine a counterfeit since the measurements are well within the minimum value of $\pm 12$~V stated in the datasheet~\cite{tl074_ti,tl074_moto,tl074_hlf}.

The simplicity of the $\mathrm{I_{CC}}$ test made it well-suited for use in undergraduate labs and informed our decision to incorporate it as a pre-lab activity in the redesigned course.

\subsection{Curriculum Adjustments and Instructional Strategy}

To address the presence of counterfeit TL074 op-amps, the lab curriculum was redesigned to incorporate a structured diagnostic process centered around the $\mathrm{I_{CC}}$ detection method. This check was introduced as a required pre-lab activity, enabling early detection of suspect ICs before students soldered them onto their PCBs.

An additional instruction document was provided alongside the lab handout to:
\begin{itemize}
    \item Explain the issue of counterfeit chips and their known behavior (e.g., reduced bandwidth, distorted signals)
    \item Provide instructions for measuring $\mathrm{I_{CC}}$ using a $1\,\Omega$  resistor method instead of a direct ammeter setup

\end{itemize}

First, TAs informed students about the presence of counterfeit ICs. As part of the pre-lab assignment, students measured $\mathrm{I_{CC}}$ on a breadboard, and consulted a TA for replacement if values were out of range. Valid measurements allowed students to proceed with soldering the IC to the PCB. If circuit results were still incorrect, the process was repeated until successful, after which students finalized their lab report. Fig.~\ref{fig:workflow} outlines this structured workflow.

\usetikzlibrary{shapes.geometric, arrows.meta, positioning}

\tikzset{
    block/.style = {
        rectangle, rounded corners, 
        minimum width=2.5cm, minimum height=0.8cm, 
        text centered, draw=black, fill=gray!10,
        font=\scriptsize, text width=2.2cm
    },
    decision/.style = {
        diamond, draw=black, fill=gray!10, 
        text centered, inner sep=0pt,
        minimum width=2cm, minimum height=0.8cm, 
        aspect=2, font=\scriptsize, text width=1.8cm
    },
    arrow/.style = {
        thin, ->, >=stealth
    }
}

\begin{figure}[htbp]
\centering
\resizebox{0.85\columnwidth}{!}{%
\begin{tikzpicture}[
    node distance=0.5cm and 0.8cm,
    every node/.style={align=center}
]

\node (start) [block] {TA announces fake ICs};
\node (prelab) [block, below=of start] {Pre-lab assignment};
\node (measure) [block, below=of prelab] {Measure $\mathrm{I_{CC}}$\\on breadboard};
\node (inrange) [decision, below=of measure] {$\mathrm{I_{CC}}$\\in range?};
\node (replace) [block, right=of inrange, xshift=1cm] {Refer to TA\\replace chip};
\node (solder) [block, below=of inrange, yshift=-0.3cm] {Solder\\to PCB};
\node (results) [decision, below=of solder] {Results\\match?};
\node (feedback) [block, below=of results] {Write report};

\draw [arrow] (start) -- (prelab);
\draw [arrow] (prelab) -- (measure);
\draw [arrow] (measure) -- (inrange);
\draw [arrow] (inrange) -- node[above] {No} (replace);
\draw [arrow] (inrange) -- node[right=2pt] {Yes} (solder);
\draw [arrow] (solder) -- (results);
\draw [arrow] (results) -- node[right] {Yes} (feedback);
\draw [arrow] (results.east) -- ++(1,0) |- node[pos=0.25, right] {No} (replace);
\draw [arrow] (replace) |- ([yshift=-1mm]measure.east);

\end{tikzpicture}
}
\caption{Student workflow for counterfeit IC detection.}
\label{fig:workflow}
\end{figure}
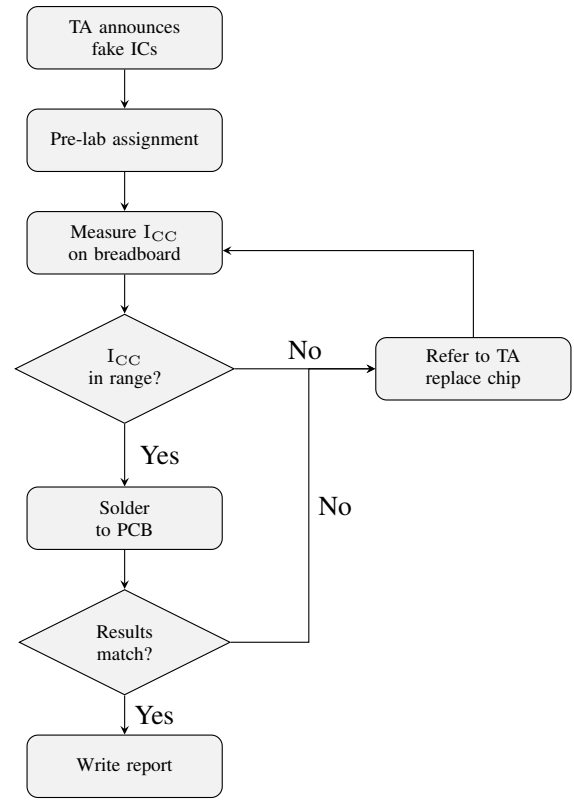

\section{Implementation Results}
This section presents implementation results drawn from TAs interviews, analyzed to extract common instructional themes.

\subsection{Data Source}
Two semi-structured interviews were conducted with TAs following the Spring 2025 offering of the course. The conversations focused on how students responded to the counterfeit op-amps, the effectiveness of the diagnostic methods used, and the instructional adaptations made during the semester. Both interviews were audio-recorded with consent and reviewed for analysis.

\subsection{Data Analysis}
The transcripts were analyzed using thematic coding based on the methodology outlined in \cite{saldana2021coding}. Key phrases and recurring observations were coded and grouped into broader themes. This process surfaced consistent patterns in TA experiences, particularly around test reliability, student engagement, and practical limitations in lab implementation.

\subsection{Themes Found}
\subsubsection{\textbf{$\mathit{\boldsymbol{I_{CC}}}$ Test Was Inconclusive in Some Cases}}

TAs reported that while the $\mathrm{I_{CC}}$ test helped identify many counterfeit ICs, it did not reliably distinguish all fake chips. Some counterfeit components passed the current test but failed during circuit operation, likely due to variation across counterfeit batches. This limitation reinforced that $\mathrm{I_{CC}}$ alone should not be used to determine authenticity.

\subsubsection{\textbf{Students Faced Challenges in Measurement and Workflow}}

Many students had difficulty accurately measuring or interpreting $\mathrm{I_{CC}}$ values. In addition, 
some students had already soldered their TL074 op-amps to the PCB before testing, making replacement more difficult. This highlighted the need for clearer pre-lab instructions and emphasized the value of testing on a breadboard first or using a 14-pin DIP socket to simplify chip replacement without desoldering.

\subsubsection{\textbf{Student Insight}}

Although the $\mathrm{I_{CC}}$ test had limitations, it sparked deeper engagement from many students. High-performing groups asked questions about circuit behavior and test reliability, which led to discussions about signal distortion and performance metrics. TAs found that combining $\mathrm{I_{CC}}$ with visual inspection, waveform inspection and frequency analysis was essential for both accurate diagnosis and fostering critical thinking.

\section{ Future Work}

The lab redesign revealed several areas for further improvement. While the $\mathrm{I_{CC}}$ test provided a valuable screening tool, its limitations showed the need to incorporate additional diagnostics—including waveform distortion analysis, frequency response measurements, and visual inspection—to improve reliability and better distinguish counterfeit components. In future offerings, visual inspection will be added to the pre-lab process, encouraging students to examine markings, logos, and package quality before performing electrical tests.

\section{Conclusion}

Counterfeit ICs have increasingly infiltrated the supply chains of even reputable vendors and distributors. The discovery of counterfeit TL074 op-amps presented an unexpected challenge in our undergraduate electronics lab. By turning this disruption into a structured learning opportunity, we enabled students to engage with real-world diagnostic techniques and think critically about component behavior. While the integration of an $\mathrm{I_{CC}}$-based detection method provided a practical starting point for identifying suspect components, it proved insufficient on its own. This limitation highlighted the need for additional diagnostics and reinforced the value of teaching students how to troubleshoot and validate circuits in uncertain, non-ideal conditions.

\Urlmuskip=0mu plus 1mu\relax

\bibliographystyle{IEEEtran}

\end{document}